\documentclass[aip,jcp,graphicx,reprint,longbibliography,noeprint,citeautoscript]{revtex4-1}

\usepackage[pdftex]{graphics}
\usepackage{graphicx}
\usepackage{amssymb,amsmath}
\usepackage{siunitx}
\usepackage{comment}
\usepackage[colorlinks=true,linkcolor=black,citecolor=black,urlcolor=black]{hyperref}

\newcommand{\ocircle}{\raisebox{0.3ex}{\scalebox{0.5}{$\bigcirc$}}}
\newcommand{\fcircle}{\ensuremath{\bullet}}
\newcommand{\osquare}{\raisebox{0.2ex}{\scalebox{0.6}{$\square$}}}
\newcommand{\fsquare}{\raisebox{0.2ex}{\scalebox{0.6}{$\blacksquare$}}}
\newcommand{\otriangle}{\raisebox{0.25ex}{\scalebox{0.6}{$\triangle$}}}
\newcommand{\ftriangle}{\raisebox{0.25ex}{\scalebox{0.65}{$\blacktriangle$}}}

\begin{document}
\preprint{AIP/123-QED}

\title{Nuclear quantum effects in the thermal conductivity of solid hydrogen\\
}

\author{Hengtai Zheng}
\affiliation{Department of Chemistry, University of Oxford, Physical and Theoretical Chemistry Laboratory, South Parks Road, Oxford, OX1 3QZ, UK}

\author{Zezhu Zeng}
\affiliation{Department of Chemistry, University of Oxford, Physical and Theoretical Chemistry Laboratory, South Parks Road, Oxford, OX1 3QZ, UK}

\author{David E. Manolopoulos}
\email{david.manolopoulos@chem.ox.ac.uk }
\affiliation{Department of Chemistry, University of Oxford, Physical and Theoretical Chemistry Laboratory, South Parks Road, Oxford, OX1 3QZ, UK}

\date{\today}

\begin{abstract}
We use a combination of path integral and lattice dynamics methods to calculate the thermal conductivity of solid parahydrogen. Path integral molecular dynamics is first used to calculate a centroid potential of mean force at each temperature, on which a harmonic phonon basis is constructed using the finite displacement method. We then calculate anharmonic force constants on the centroid potential and solve the linearised phonon Boltzmann transport equation for the thermal conductivity. The resulting renormalised phonon dispersion curves, heat capacity, and thermal conductivity are all in remarkably good agreement with experimental measurements. We find that the heat transport is dominated by collective relaxon modes at low temperatures and cannot be captured by making the phonon relaxation time approximation. Indeed, quantitative agreement with the experimental thermal conductivity is only achieved when the nuclear quantum effects in the centroid potential of mean force, phonon boundary scattering, and hydrodynamic transport are all explicitly considered. This is also likely to be the case for other quantum crystals.
\end{abstract}

\maketitle

\section{\label{sec:level1}Introduction}

Solid parahydrogen (p-H$_2$) is an unusually clean example of a molecular crystal that exhibits non-degenerate quantum heat transport. The small mass of the hydrogen molecule leads to significant intermolecular zero-point energy at all temperatures below the triple point ($T_{\rm tp}=13.8$ K), and therefore throughout the solid state. This modifies the equilibrium structure and, more importantly for heat transport, the effective potential on which the lattice vibrations occur. The lattice vibrations themselves are quantised and their thermal excitation is governed by Bose--Einstein statistics. A reliable description of the thermal conductivity must therefore account for both the nuclear quantum effects that renormalise the lattice dynamics and the quantum statistics of the phonons that carry the heat.

There are several features of p-H$_2$ that facilitate the modelling of these two effects. Since the rotational temperature of a hydrogen molecule ($\theta_{\rm rot}=85.4$ K) is well above the triple point temperature, the overwhelming majority of p-H$_2$ molecules are in their ground rotational state throughout the solid state. The resulting spherical symmetry allows the intermolecular interactions to be described by an isotropic pair potential that includes an effective pairwise representation of the leading three-body dispersion interactions.\cite{Silvera1978} The hydrogen molecules are sufficiently localised on their lattice sites that bosonic exchange can be neglected, thereby simplifying the path integral treatment of nuclear quantum effects, and the solid has a hexagonal close-packed (hcp) crystal structure that is ideal for lattice dynamics calculations. Last but not least, high quality experimental measurements of the phonon dispersion relations,\cite{Nielsen1973} heat capacity,\cite{Ahlers1964} and thermal conductivity\cite{Hill1958} of solid p-H$_2$ have been available for many years for comparison with theoretical predictions. It is hard to think of another molecular crystal with features that combine so nicely to make both the required calculations and their validation so convenient.

Many of these simplifications also apply to liquid p-H$_2$, which has therefore become a standard model of a non-degenerate quantum liquid. There have been numerous applications of imaginary time path integral methods to this liquid,\cite{Calhoun1996,Kinugawa1998,Reichman2001,Rabani2002,Miller2005,Craig2006} but only two we are aware of that have considered its thermal conductivity. 
An early Green--Kubo centroid molecular dynamics\cite{Cao1994b} (CMD) calculation by Yonetani and Kinugawa\cite{Yonetani2004} found the thermal conductivity $\kappa$ to increase monotonically with decreasing temperature, at variance with the experimental curve which has a maximum at 22 K.\cite{Roder1970} Sutherland {\em et al.}\cite{Sutherland2021} subsequently showed that this maximum can be recovered by combining the CMD thermal diffusivity $a$ with the liquid density $\rho$ and the quantum mechanical heat capacity $C_v$ to give $\kappa=\rho a C_v$. 
The reason for this is instructive: CMD provides a good description of the density fluctuations that determine \(a\),\cite{Sutherland2021} but its classical dynamics is missing the quantum energy statistics that determine $C_v$. In the solid these same energy statistics appear explicitly in the Bose--Einstein occupation numbers of the phonons. These have a profoundly non-classical effect on the phonon scattering at low temperatures that could not be captured simply by scaling the Green-Kubo CMD result to give the correct heat capacity.
 
The approach we shall take here is to follow the liquid state work by using a path integral method to fold the equilibrium nuclear quantum effects into a centroid potential of mean force (CPMF), but then perform quantised lattice dynamics on this potential rather than classical molecular dynamics. Having decided to do this, the only remaining question is what flavour of lattice dynamics to use. The phonon relaxation time approximation (RTA) is a popular choice because it only requires the diagonal matrix elements of the phonon collision matrix and it is expected to work well when Umklapp processes are strong enough to relax the phonon crystal momentum. However, at the lowest temperatures for which experimental results are available, we shall find that Umklapp processes are suppressed, normal scattering dominates, the heat transport becomes hydrodynamic,\cite{Guyer1966b} and the RTA breaks down. We shall therefore use the full solution of the linearised Boltzmann transport equation (BTE) for our lattice dynamics calculations,\cite{Omini1996} and include the RTA results to illustrate their breakdown. 

The literature on thermal transport in solid p-H$_2$ dates back to the 1950s, but we are not aware of any previous study that has calculated its intrinsic (perfect crystal) phonon scattering by solving the full BTE, or of any study that has first renormalised the phonon basis to include nuclear quantum effects by calculating it on the CPMF. Most of the earlier theoretical investigations have either used the RTA or phenomenological Callaway-type\cite{Callaway1959} treatments of normal and Umklapp scattering, with a particular focus on understanding the extrinsic scattering by orthohydrogen (o-H$_2$) impurities.\cite{Ebner1970,Kokshenev1975} We shall deliberately compare our computed thermal conductivity with an experiment on a sample with a low (0.5\%) measured o-H$_2$ concentration to minimise the effect of this scattering, and bear what may remain of it in mind when we come on to discuss our results.

\section{Centroid potential of mean force}

\def\p{\boldsymbol{p}}
\def\r{\boldsymbol{r}}
\def\f{\boldsymbol{f}}

\subsection{Theory}

The CPMF at each NVT state point folds the equilibrium nuclear quantum effects in the partition function of the state point into an effective classical interaction potential $V^{(c)}(\r)$.\cite{Feynman1965} Above the melting temperature of p-H$_2$, these nuclear quantum effects are already strong enough to change a state point in the classical liquid-gas coexistence region\cite{Miller2005b} into a point in the quantum liquid state.\cite{Miller2005} Since they are no less pronounced at the lower temperatures of the solid, we shall find it essential to do our lattice dynamics calculations on $V^{(c)}(\r)$  rather than on the bare classical interaction potential $V(\r)$.

The partition function $Z={\rm tr}[e^{-\beta\hat{H}}]$ can be written as the limit as ${P\to\infty}$ of a discretised path integral 
\begin{equation}
Z_P = \frac{1}{(2\pi\hbar)^{3NP}}\int\int \prod_{p=1}^{P} {\rm d}\p^{(p)}{\rm d}\r^{(p)}\,e^{-\beta_PH_P({\bf p},{\bf r})},
\end{equation}
where $N$ is the number of p-H$_2$ molecules, $P$ is the number of path integral beads, and $\beta_P=\beta/P$ with $\beta=1/k_{\rm B}T$. Here $H_P({\bf p},{\bf r})$ is a classical ring polymer Hamiltonian\cite{Chandler1981,Parrinello1984,Craig2004}
\begin{align}
H_P({\bf p},{\bf r}) &=
\sum_{p=1}^{P} \left[
\frac{|\p^{(p)}|^2}{2m}+\frac{1}{2}m\omega_P^2|\r^{(p)}-\r^{(p-1)}|^2+V(\r^{(p)})\right],\nonumber\\
\end{align}
in which $m$ is the mass of a hydrogen molecule,  $\omega_P=1/\beta_P\hbar$, and $\r^{(0)}\equiv \r^{(P)}$.

The CPMF is defined to within an additive constant in this notation by
\begin{equation}
V^{(c)}(\r) = -\frac{1}{\beta}\ln \left<\frac{\delta(\r-\r^{(c)})}{\rho_0}\right>,
\end{equation}
where $\rho_0$ is an arbitrary reference density with the same dimensions as $\delta(\r-\r^{(c)})$, $\r^{(c)}$ is the centroid of the ring polymer
\begin{equation}
\r^{(c)} = \frac{1}{P}\sum_{p=1}^{P} \r^{(p)},
\end{equation}
and the angular brackets denote the path integral average
\begin{equation}
\left<\cdots\right> = \frac{1}{(2\pi\hbar)^{3NP}Z_P}\int\int \prod_{p=1}^{P} {\rm d}\p^{(p)}{\rm d}\r^{(p)}\,e^{-\beta_PH_P({\bf p},{\bf r})}(\cdots).
\end{equation} 
Differentiating Eq.~(3) with respect to $\r$ and
integrating once by parts gives the associated centroid force as
\begin{equation}
\f^{(c)}(\r) = -\frac{\partial V^{(c)}(\r)}{\partial \r} = -\left< \frac{1}{P}\sum_{p=1}^{P} \frac{\partial V(\r^{(p)})}{\partial \r^{(p)}}\right>_{\r^{(c)}=\r},
\label{fc}
\end{equation}
where $\left<\cdots\right>_{\r^{(c)}=\r}$ is the centroid-constrained average
\begin{equation}
\left<\cdots\right>_{\r^{(c)}=\r} = \frac{\left<\delta(\r-\r^{(c)})(\cdots)\right>}{\left<\delta(\r-\r^{(c)})\right>}.
\label{cavg}
\end{equation}

Most methods for calculating $V^{(c)}(\r)$ involve fitting the forces $\f^{(c)}(\r)$ obtained from a path integral simulation to those of a suitable model potential -- such as a neural network potential\cite{Musil2022} or a pairwise correction to the classical interaction potential\cite{Hone2005} -- by non-linear least squares. A similar approach is taken in the quantum version of the temperature-dependent effective potential (TDEP) method,\cite{Hellman2013} which least-squares fits the forces  $\f^{(c)}(\r)$ to a Taylor expansion of $V^{(c)}(\r)$ around the equilibrium geometry $\r_0=\langle \r^{(c)}\rangle$ of the path integral simulation to obtain second- and higher-order force constants for use in anharmonic lattice dynamics calculations. For the present simulations of solid hydrogen we have found it simpler to follow Sutherland {\em et al.}\cite{Sutherland2021} and exploit the fact that the Silvera-Goldman pair potential\cite{Silvera1978} already provides an excellent description of the interactions in a system of rotationless $(J=0)$ hydrogen molecules. Adding a pairwise correction to this results in a pairwise approximation to $V^{(c)}(r)$ that can be extracted directly from the path integral simulation.\cite{Sutherland2021} 

Suppose we write the classical interaction potential as 
\begin{equation}
V(\r) = \frac{1}{2}\sum_{i=1}^N\sum_{j\not=i}^N v(r_{ij}),
\end{equation}
where $v(r)$ is the Silvera-Goldman pair potential and $r_{ij}=|\r_{ij}|$ with $\r_{ij}=\r_i-\r_j$, and we approximate the CPMF as
\begin{equation}
V^{(c)}(\r) \approx \frac{1}{2}\sum_{i=1}^N\sum_{j\not=i}^N v^{(c)}(r_{ij}).
\end{equation}
Then since the radial component of the centroid force between molecules $i$ and $j$ at a configuration ${\bf r}$ of a path integral simulation is
\begin{equation}
f_{ij}^{(c)}({\bf r}) = -\frac{1}{P} \sum_{p=1}^P \hat{\r}_{ij}^{(c)}\cdot\hat{\r}_{ij}^{(p)}v'(r_{ij}^{(p)}),
\end{equation}
where $\hat{\r}_{ij}^{(c)}$ and $\hat{\r}_{ij}^{(p)}$ are unit vectors in the directions $\r_{ij}^{(c)}=\r_i^{(c)}-r_j^{(c)}$ and $\r_{ij}^{(p)}=\r_i^{(p)}-r_j^{(p)}$, we can compute $v^{(c)}(r)$ as\cite{Sutherland2021}
\begin{equation}
v^{(c)}(r) = v^{(c)}(r_{\rm max})+\int_r^{r_{\rm max}} 
f^{(c)}(s)
\,{\rm d}s
\end{equation}
where
\begin{equation}
    f^{(c)}(s) = \left<f_{ij}^{(c)}({\bf r})\right>_{r_{ij}^{(c)}=s}.
\end{equation}

The conditional average on the right hand side of Eq.~(12) can be estimated by binning $s$ and updating $f^{(c)}(s)$ in each bin that contains an $r_{ij}^{(c)}$ at each configuration of an unconstrained path integral molecular dynamics (PIMD) simulation. The horizontal statistics provided by the $N(N-1)/2$ values of $r_{ij}^{(c)}$ at each path integral configuration ${\bf r}$ improve the convergence allowing the use of relatively narrow histogram bins. The integral in Eq.~(11) can be evaluated using the mid-point rule, with $r_{\rm max}$ set equal to the cutoff radius of the Silvera-Goldman pair potential and $v^{(c)}(r_{\rm max})$ set equal to zero (thereby eliminating the arbitrary constant shift in $V^{(c)}(\r)$ that comes from the $\rho_0$ in Eq.~(3)). The resulting pair potential can then be least-squares fit over the range of centroid distances accessed in the path integral simulation to a smooth function of the form
\begin{equation}
v^{(c)}(r) = \sum_{k=0}^m c_k\left[\left(\frac{r_{\rm max}}{r}\right)^{k+6}-1\right]
\end{equation}
for use in lattice dynamics calculations. We used $r_{\rm max}=12\,\mathring{\rm A}$ and $m=9$ in this equation for all of the calculations in this paper.

\subsection{Computational details}

The experimentally determined lattice parameters of solid p-H$_2$ change by less than 0.5\% between 2 K and the melting point,\cite{Krupskii1983} so we used the same average hcp lattice parameters $a=3.79\,\mathring{\rm A}$ and $c=\sqrt{8/3}\,a$ in all of our calculations. A 448 molecule orthorhombic supercell was constructed from this primitive unit cell for our PIMD simulations, in which we first equilibrated the system for 10 ps at each temperature before running a further 1 ns of NVT dynamics with a time step of 2 fs to accumulate centroid forces. The temperature was controlled by applying a path integral Langevin equation thermostat\cite{Ceriotti2010} to each ring polymer internal mode and a stochastic velocity rescaling thermostat\cite{Bussi2007} to the centroid. The centroid forces were used to calculate PIMD centroid radial distribution functions (RDFs) with the force sampling method\cite{Borgis2013,Rotenberg2020} and centroid pair potentials $v^{(c)}(r)$ from Eq.~(11). We found that $P=128$ economised path integral beads (Eco\cite{Zeng2026} beads with $\hbar\omega_{\rm max}=12$ meV obtained from the top of the experimental phonon dispersion) sufficed to converge these quantities at 2.5 K, 64 at 5 K, and 32 at 10 K. To check that our code was working correctly, we repeated the calculation of the CPMF with $P=1$, and found that the resulting $v^{(c)}(r)$ agreed with the classical pair potential $v(r)$ to graphical accuracy. 

\begin{figure}[b]
    \centering
    \resizebox{0.85\columnwidth}{!}{\includegraphics{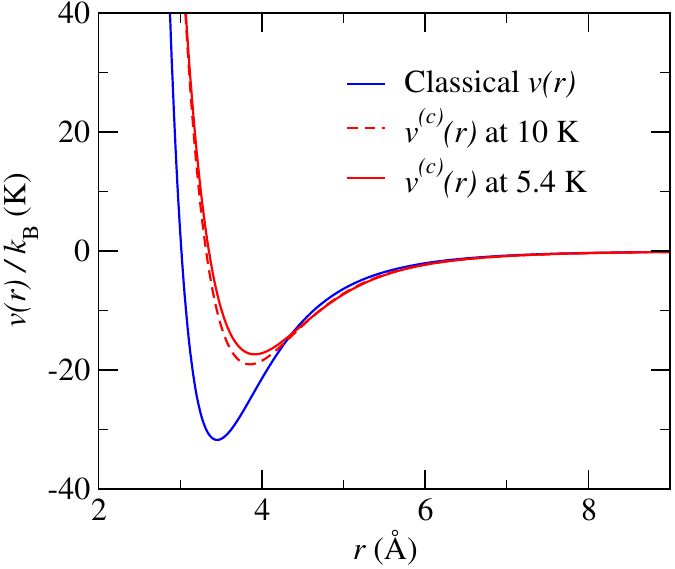}}
    \caption{The classical pair potential $v(r)$ and the centroid pair potentials of mean force $v^{(c)}(r)$ of solid p-H$_2$ at 10 K and 5.4 K.}
    \label{fig:1}
\end{figure}

\begin{figure}[htbp]
    \centering
    \resizebox{0.85\columnwidth}{!}{\includegraphics{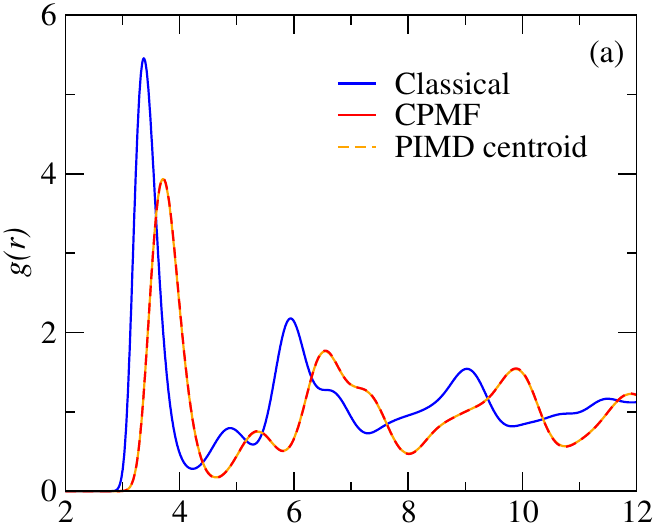}}    
    \vspace{1em} 
    \resizebox{0.85\columnwidth}{!}{\includegraphics{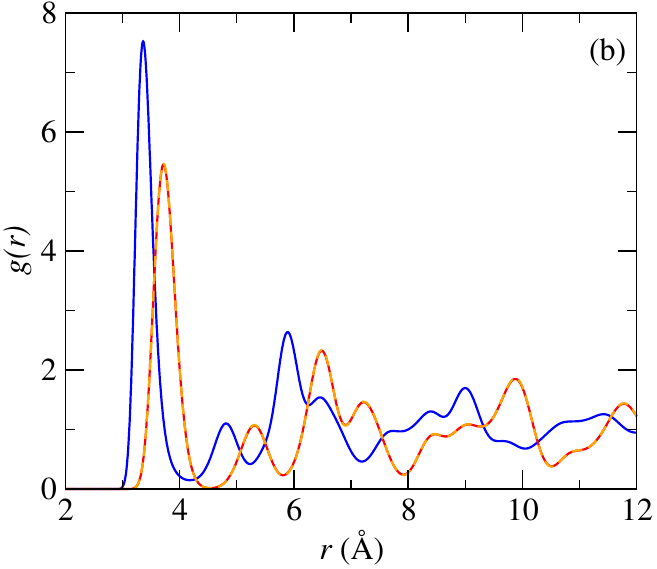}}
    \caption{Classical, CPMF, and PIMD centroid RDFs of solid p-H$_2$ at (a) 10 K and (b) 5.4 K.}
    \label{fig:2}
\end{figure}

\subsection{Results and discussion}

Fig.~\ref{fig:1} compares the centroid pair potentials of mean force at $T=10$ K and $5.4$ K (the temperature we shall use to compare with the experimental phonon dispersion in Sec.~III) with the underlying Silvera-Goldman pair potential. The centroid potentials are more repulsive at short range and more attractive at long range than the classical potential because the finite bead delocalisation samples the strongly convex repulsive wall at short range, enhancing the effective repulsion, while at longer range the inequality $\langle r^{-n}\rangle \ge \langle r\rangle^{-n}$ enhances the attractive inverse-power contributions. The increase in the short-range repulsion dominates and leads to a shallower potential well. This can be interpreted as a zero-point energy effect: the centroid potentials of mean force include a contribution from the zero-point energy in the well whereas the classical potential does not. Sutherland {\em et al.}\cite{Sutherland2021} have plotted similar centroid pair potentials for liquid p-H$_2$ at $T=25$ K and 14 K. The only difference between their results and ours is that the departure from the classical potential is more pronounced at our lower temperatures.  The classical potential can be regarded as the high-temperature limit of the CPMF in which the ring polymer has collapsed onto its centroid and the zero-point energy has become negligible compared with $k_{\rm B}T$.

Fig.~\ref{fig:2} compares the classical, CPMF, and PIMD centroid RDFs of solid p-H$_2$ at the same two temperatures. The classical and CPMF RDFs were computed from classical NVT trajectories on the potentials $V(\r)$ and $V^{(c)}(\r)$, and the PIMD centroid RDF was calculated during the PIMD simulation as described in Sec.~II.A. Sutherland {\em et al.}\cite{Sutherland2021} have presented the analogous plots for liquid p-H$_2$ at $T=25$ K and 14 K. Their plots are missing the long-range order we see in the solid, but they are similar in all other respects. The nuclear quantum effects in both phases shift the RDF to larger $r$ because they strengthen the short-range repulsion, and they broaden the first peak in the RDF because they reduce the depth and the curvature of the pair potential well. The fact that the CPMF and PIMD centroid RDFs in Fig.~\ref{fig:2} are the same to graphical accuracy implies that the method outlined in Sec.~II.A is finding the unique effective classical pair potential that is consistent with the PIMD centroid RDF to within an additive constant,\cite{Henderson1974} which we have removed by setting the large $r$ limit of $v^{(c)}(r)$ to zero.

\section{Harmonic Phonon Basis}

\def\R{\boldsymbol{R}}
\def\q{\boldsymbol{q}}
\def\t{\boldsymbol{\tau}}

\subsection{Theory}

There are two different ways to construct a harmonic phonon basis on the CPMF $V^{(c)}(\r)$. One can either use the forces $\f^{(c)}(\r)$ calculated at small displacements from the equilibrium geometry $\r_0$ of the potential to calculate a finite-difference approximation to the Hessian as is done in the finite displacement method,\cite{Parlinski1997} or one can perform a classical molecular dynamics simulation on the CPMF and construct an effective Hessian from the resulting displacements as is done in the TDEP method.\cite{Hellman2013} We have explored both approaches and found that using the Hessian of the CPMF at its equilibrium geometry gives slightly better agreement with the experimental phonon dispersion curves of solid p-H$_2$.\cite{Nielsen1973} Since this is also the simpler of the two methods, we have used it in all of our calculations.

Thus we define our Hessian to be the second-order force constant matrix of the CPMF,
\begin{equation}
\Phi_{lb\alpha,l'b'\alpha'}=\frac{\partial^2 V^{(c)}(\r)}
{\partial r_{lb\alpha}\partial r_{l'b'\alpha'}}
\Biggr|_{\boldsymbol{\r}=\boldsymbol{\r_0}},
\end{equation}
where $r_{lb\alpha}$ is one of the Cartesian components of the position of the $b$-th molecule in the $l$-th unit cell of the crystal and $\r_0$ is the equilibrium geometry of the crystal lattice (the position of the global minimum of $V^{(c)}(\r)$). The corresponding harmonic phonon basis is the eigenbasis of the dynamical matrix at each wavevector $\q$,
\begin{equation}
D_{b\alpha,b'\alpha'}(\q)=\frac{1}{\sqrt{m_bm_{b'}}}\sum_{l'}\Phi_{0b\alpha,l'b'\alpha'}
e^{+i\q\cdot(\R_{l'}+\t_{b'}-\t_b)},
\end{equation}
where $\t_b$ is the equilibrium position of the $b$-th molecule in the $l$-th unit cell relative to the lattice vector $\R_l$ and in our case $m_b=m_{b'}=m$ (the mass of a hydrogen molecule). The solution of the eigenvalue problem 
\begin{equation}
D(\q)\boldsymbol{\epsilon}_s(\q)=
\omega_{\q s}^2\boldsymbol{\epsilon}_s(\q),
\end{equation}
yields the harmonic phonon frequencies $\omega_{\q s}$ and polarisation vectors $\boldsymbol{\epsilon}_s(\q)$, from which the harmonic normal mode displacements can be calculated as
\begin{equation}
u_{\q s} = \frac{1}{\sqrt{N_{\q}}}\sum_{lb\alpha}\sqrt{m_b}\,u_{lb\alpha}\,  e^{-i\q\cdot(\R_l+\t_b)}\epsilon_{s, b\alpha}^*(\q),
\end{equation}
where $N_{\q}$ is the number of $\q$ points in a mesh commensurate with the supercell and $u_{lb\alpha}$ is the displacement of $r_{lb\alpha}$ from its equilibrium value.

\subsection{Computational details}

We began by using a homemade code to calculate the second-order force constants on the CPMF from finite displacements,  using a $7\times 4\times 4$ orthorhombic supercell containing 448 p-H$_2$ molecules, a finite difference displacement of $2^{-16}\,a_0$, and a force cutoff distance of $12\,\mathring{\rm A}$. Because each Hessian matrix element was obtained by finite-differencing analytical forces, the resulting Hessian was explicitly symmetrised before being used to assemble the dynamical matrix at each $\q$ point. As a check, we repeated the calculation using the open-source package \texttt{Phonopy},\cite{Togo2015} with a $6\times 6\times 3$ hcp supercell containing 216 p-H$_2$ molecules, a (default) displacement of $0.01\,\mathring{\rm A}$, and a force cutoff distance of $9\,\mathring{\rm A}$. The dispersion curves calculated from the two codes were found to agree to graphical accuracy.

\subsection{Results and discussion}

\begin{figure}[htbp]
    \centering

    \resizebox{0.85\columnwidth}{!}{\includegraphics{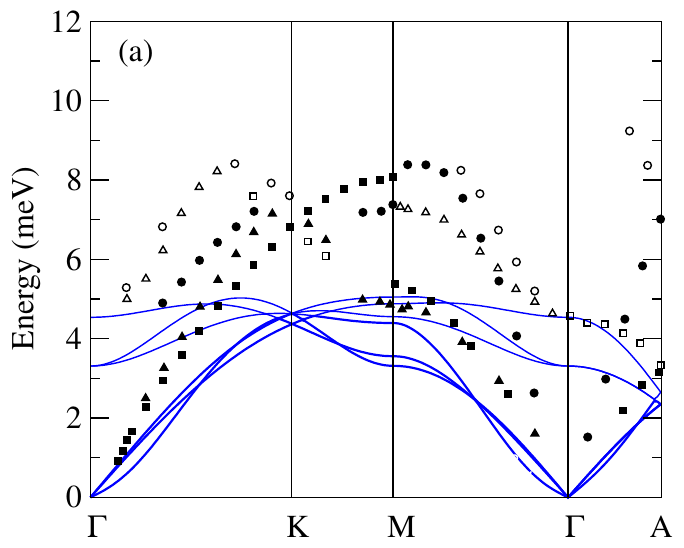}}  
    \resizebox{0.85\columnwidth}{!}{\includegraphics{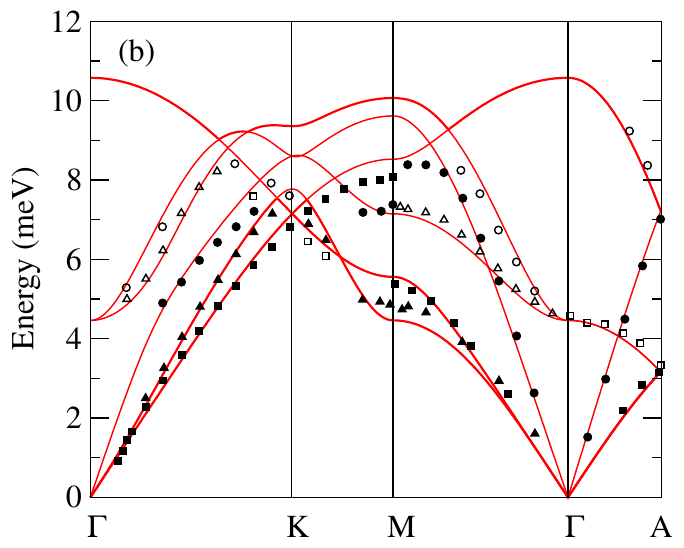}} 

    \caption{Comparison of the harmonic phonon dispersions calculated on (a) $V(\r)$ (blue curves) and (b) $V^{(c)}(\r)$ (red curves) with the experimental p-H$_2$ neutron scattering data of Nielsen\cite{Nielsen1973} (black symbols). Both the measurements and the calculations were done at 5.4 K. The high symmetry points are those of the first Brillouin zone of the hcp lattice. The phonon branches are $\fcircle$ LA, $\ocircle$ LO, $\fsquare$ T$_{\perp}$A, $\osquare$ T$_{\perp}$O, $\ftriangle$ T$_{\parallel}$A, and $\otriangle$ T$_{\parallel}$O.}
    
    \label{fig:3}
\end{figure}

The harmonic phonon dispersion curves on the CPMF at 5.4 K are compared with the neutron scattering data of Nielsen\cite{Nielsen1973} in Fig.~\ref{fig:3}. We have also included the results obtained by replacing $V^{(c)}(\r)$ in Eq.~(14) with the classical interaction potential $V(\r)$ to shed light on the role of nuclear quantum effects. Since the phonon frequencies computed on $V(\r)$ are temperature-independent, whereas those computed on $V^{(c)}(\r)$ depend on the temperature and Planck's constant, one could describe the former as harmonic and the latter as renormalised or quasi-harmonic. We shall use the term ``renormalised" in what follows to avoid confusion with other quasi-harmonic approximations in the literature,\cite{Allen2020} in which the harmonic frequencies only depend on temperature implicitly through the volume at each temperature $V(T)$.

The phonon dispersion curves obtained on $V^{(c)}(\r)$ are in remarkably good agreement with the neutron scattering data, whereas the results obtained on $V(\r)$ are not. The increased repulsion in the centroid pair potential hardens the solid and roughly doubles the phonon frequencies, bringing them closer to the experiment. The remaining discrepancies between the CPMF and experimental dispersion curves are most visible in the optical branches. Since these have a negligible population at 5.4 K, where $k_{\rm B}T$ is less than 0.5 meV, these discrepancies will have a negligible impact on properties such as the heat capacity and the thermal conductivity of the solid at this temperature. 

The only previous theoretical study of these phonon dispersion curves we are aware of was performed by Saito {\em et al.},\cite{Saito2003} who used CMD to calculate dynamic structure factors $S(\q,\omega)$ at particular $\q$ points and inferred the phonon frequencies from the positions of their peaks. Since CMD includes classical thermal excitations on the CPMF, this approach is analogous to using the TDEP method to calculate the Hessian on $V^{(c)}(\r)$. 
Saito {\em et al.}'s dispersion curves are in less good agreement with Nielsen's experiment\cite{Nielsen1973} than ours simply because they used a different density for their calculations. For example, their longitudinal acoustic (LA) phonon has an energy above 11 meV at the M point, and their longitudinal optical (LO) phonon has an energy above 12 meV at the $\Gamma$ point.\cite{Saito2003} 

Saito {\em et al.} began by using their implementation of CMD to calculate the molar volume of the crystal at zero pressure, for which they obtained $V=21.9$ cm$^{3}$/mol.\cite{Saito2003} Since this is smaller than the experimental zero-pressure molar volume of 23.1 cm$^3$/mol at 5.4 K,\cite{Krupskii1983} and also smaller than the average solid p-H$_2$ molar volume of 23.2 cm$^3$/mol we have used throughout our calculations, Saito {\em et al.}'s solid was more compressed than ours resulting in higher phonon frequencies. We have checked this by repeating our CPMF calculations with their molar volume and found much better agreement with their phonon dispersion curves. A few far smaller discrepancies remain, but we suspect they can be attributed to their use of CMD to calculate the phonon frequencies from the peaks in $S(\q,\omega)$. We used the equilibrium harmonic force constants of the CPMF.

\begin{figure}[t]
    \centering
    \resizebox{0.85\columnwidth}{!}{\includegraphics{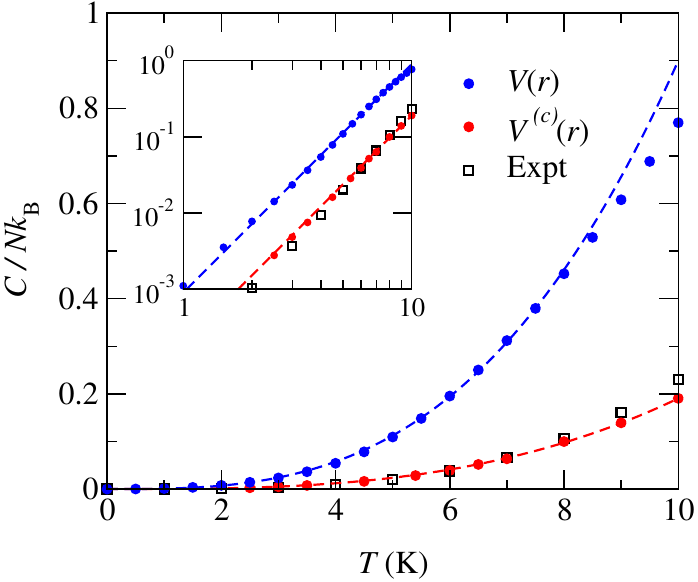}}  

    \caption{Harmonic heat capacities of solid p-H$_2$ calculated using the classical potential, and renormalised heat capacities calculated using the CPMF, compared with the zero pressure experimental data of Ahlers.\cite{Ahlers1964} The dashed lines are Debye $T^3$ fits to the theoretical results.}
    
    \label{fig:4}
\end{figure}

The harmonic phonon frequencies on the classical potential and the CPMF can also be used to calculate the heat capacities
\begin{equation}
    C =\sum_{\lambda} C_{\lambda},
\end{equation}
where
\begin{equation}
C_{\lambda} = \hbar\omega_{\lambda}
\left(\frac{\partial \bar{n}_{\lambda}}{\partial T}\right)_{\omega_\lambda}
= k_B
\left(\frac{\hbar\omega_{\lambda}}{k_BT}\right)^2
\bar{n}_{\lambda}(\bar{n}_{\lambda}+1)
\end{equation}
with $\lambda=(\q, s)$ and $\bar{n}_{\lambda} = (e^{\hbar\omega_{\lambda}/k_BT}-1)^{-1}$. We have used \texttt{Phonopy}\cite{Togo2015} with a converged $25\times 25\times 15$ hcp $\q$-mesh to compute these heat capacities at temperatures up to 10 K. In the case of the CPMF calculation the renormalised heat capacity at each temperature was evaluated using Eq.~(19) with the harmonic phonon frequencies of the CPMF constructed at that temperature. The results of both calculations are compared with the experimental zero pressure heat capacity of solid p-H$_2$\cite{Ahlers1964} in Fig.~4.

At these low temperatures the heat capacity is dominated by the contribution from the acoustic phonons and is almost perfectly Debye-like, as can be seen from the $C\propto T^3$ fits to the theoretical data. The heat capacity calculated on $V(\r)$ is well above the experimental result because the harmonic frequencies of this potential are too low (see Fig.~3a), whereas the heat capacity calculated on $V^{(c)}(\r)$ is in remarkably good agreement with the experiment. We suspect the main source of the remaining discrepancy is that we have used a constant molar volume of 23.2 cm$^3$ in our calculations, whereas the experimental measurements were performed at zero pressure.\cite{Ahlers1964} The slight increase in the experimental molar volume with increasing $T$ softens the phonons, lowers their frequencies, and increases the heat capacity. As a result, a fit of the experimental $C_p$ data to $AT^{a}$ gives an exponent of 3.4, whereas the exponent of our numerical CPMF $C_v$ data is 3.0.

Since these comparisons of phonon dispersion relations and heat capacities with experiment show that the CPMF $V^{(c)}(\r)$ provides a much better starting point for thermal conductivity calculations than the classical potential $V(\r)$, we shall abandon the classical potential from this point on. 

\section{Anharmonic Lattice Dynamics}

\def\k{\boldsymbol{\kappa}}
\def\J{\mathbf{J}}
\def\v{\mathbf{v}}
\def\V{\mathbf{V}}

\subsection{Theory}

The thermal conductivity tensor $\boldsymbol{\kappa}$ is defined through Fourier's law
\begin{equation}
\J = -\k\nabla T,
\end{equation}
where $\nabla T$ is a small steady-state temperature gradient and $\mathbf{J}$ is the resulting heat flux density. The diagonal harmonic phonon expression for $\J$ is\cite{Hardy1963}
\begin{equation}
\J = \frac{1}{V}\sum_{\lambda} \hbar\omega_{\lambda}\v_{\lambda}\Delta n_{\lambda},
\end{equation}
where $V$ is the volume of the crystal, $\v_{\lambda}$ is the phonon group velocity $\v_{\q s}=\partial \omega_{\q s}/\partial \q$, and $\Delta n_{\lambda}$ is the departure of the phonon occupation number from the local Bose--Einstein equilibrium distribution,
\begin{equation}
\Delta n_{\lambda}(\r) = n_{\lambda}(\r)-\bar{n}_{\lambda}(T(\r)).
\end{equation}
This can be obtained by solving the steady-state version of the linearised Boltzmann transport equation (BTE)
\begin{equation}
    \sum_{\lambda'} \Omega_{\lambda\lambda'}\Delta n_{\lambda'} = -\frac{\partial \bar{n}_{\lambda}}{\partial T}\v_{\lambda}^{\sf T}\nabla T,
\end{equation}
in which $\Omega_{\lambda\lambda'}$ is an element of a collision matrix that accounts for phonon scattering. 

The problem is usually symmetrized before solving Eq.~(23) by defining a similarity transformed collision matrix with elements
\begin{equation}
\tilde{\Omega}_{\lambda\lambda'} = \Omega_{\lambda\lambda'}\sqrt{\frac{\bar{n}_{\lambda'}(\bar{n}_{\lambda'}+1)}{\bar{n}_{\lambda}(\bar{n}_{\lambda}+1)}},
\end{equation}
which is real and symmetric by detailed balance and positive semi-definite because scattering leads to non-negative entropy production.\cite{Ziman2001} One also defines
\begin{equation}
    \Delta\tilde{n}_{\lambda} = \frac{\Delta n_{\lambda}}{\sqrt{\bar{n}_{\lambda}(\bar{n}_{\lambda}+1)}}
\end{equation}
and 
\begin{equation}
\theta^0_{\lambda} = \sqrt{\frac{C_{\lambda}}{C}},
\end{equation}
so that Eq.~(21) becomes
\begin{equation}
\J = \frac{\sqrt{k_{\rm B}T^2C\vphantom{|}}}{V}\sum_{\lambda} \theta_{\lambda}^0\v_{\lambda}\Delta \tilde{n}_{\lambda},
\end{equation}
and Eq.~(23) becomes
\begin{equation}
    \sum_{\lambda'} \tilde{\Omega}_{\lambda\lambda'}\Delta \tilde{n}_{\lambda'} = -\sqrt{\frac{C}{k_{\rm B}T^2}}\,\theta^0_{\lambda}\v_{\lambda}^{\sf T}\nabla T.
\end{equation}
Solving Eq.~(28) for $\Delta \tilde{n}_{\lambda}$, substituting the result into Eq.~(27), and comparing with Fourier's law gives 
\begin{equation}
    \k = \frac{C}{V} \sum_{\lambda\lambda'} \theta_{\lambda}^0\v_{\lambda}\tilde{\Omega}^{+}_{\lambda\lambda'}\v_{\lambda'}^{\sf T}\theta_{\lambda'}^0,
\end{equation}
where ${\tilde{\Omega}}^+$ is the pseudo-inverse of the matrix ${\tilde{\Omega}}$. This can be written more compactly in Dirac notation as
\begin{equation}
    \k  = \frac{C}{V}\langle 0|\hat{\v}\,\hat{\tilde{\Omega}}^+\hat{\v}^{\sf T}|0\rangle,
\end{equation}
where  $\hat{\v}$ is the group velocity operator
\begin{equation}
\hat{\v} = \sum_{\lambda} |\lambda\rangle\v_{\lambda}\langle\lambda|,
\end{equation}
$\hat{\tilde{\Omega}}^+$ is the pseudo-inverse of the symmetrised collision operator
\begin{equation}
\hat{\tilde{\Omega}} = \sum_{\lambda\lambda'} |\lambda\rangle \tilde{\Omega}_{\lambda\lambda'} \langle\lambda'|,
\end{equation}
and $\langle 0|\lambda\rangle=\langle \lambda|0\rangle=\theta_{\lambda}^0$. The need for the pseudo-inverse arises because $\hat{\tilde{\Omega}}$ is positive semi-definite, and indeed if all of the included scattering processes are energy-conserving the state $|0\rangle$ is itself an eigenstate of $\hat{\tilde{\Omega}}$ with eigenvalue 0.\cite{Guyer1966}

The pseudo-inverse can be constructed from the eigendecomposition of $\hat{\tilde{\Omega}}$,
\begin{equation}
\hat{\tilde{\Omega}} = \sum_{\alpha} |\alpha\rangle \gamma_{\alpha} \langle\alpha |, 
\end{equation}
as
\begin{equation}
\hat{\tilde{\Omega}}^+ = \sideset{}{'}\sum_{\alpha} |\alpha\rangle \tau_{\alpha} \langle\alpha |,
\end{equation}
where $\tau_{\alpha}=1/\gamma_{\alpha}$ and the prime on the sum excludes the states for which $\gamma_{\alpha}=0$. Substituting this into Eq.~(30) gives a kinetic theory-like expression for $\k$,\cite{Cepellotti2016} 
\begin{equation}
\k = \frac{1}{V}\sideset{}{'}\sum_{\alpha} C\,\V_{\alpha}\V_{\alpha}^{\sf T}\tau_{\alpha},
\end{equation}
where $\V_{\alpha} = \langle 0|\hat{\v}|\alpha\rangle = \langle \alpha |\hat{\v}|0\rangle$. This is the {\em relaxon} picture in which the heat is transported by collective relaxons (linear combinations of phonons) with well-defined lifetimes $\tau_{\alpha}$ and mean free path vectors $\boldsymbol{\Lambda}_{\alpha}=\V_{\alpha}\tau_{\alpha}$, as is discussed in more detail in the paper by Cepellotti and Marzari.\cite{Cepellotti2016}

One cannot obtain such an intuitive picture of the heat transport in the phonon basis without making an approximation because the collision operator is not diagonal in the phonon basis. The well-known phonon relaxation time approximation (RTA) simply makes it diagonal by discarding its off-diagonal phonon matrix elements.\cite{Omini1996,Fugallo2013} This leads to another kinetic theory-like expression in which the heat is transported by phonons,
\begin{equation}
\k^{\rm RTA} = \frac{1}{V}\sum_{\lambda} C_{\lambda} \v_{\lambda} \v_{\lambda}^{\sf T} \tau^{\rm RTA}_{\lambda},
\end{equation}
with $\tau^{\rm RTA}_{\lambda}={\Omega}_{\lambda\lambda}^{-1}$. Despite its widespread use this is ultimately just an uncontrolled approximation to the full BTE expression for $\k$ in Eq.~(35). There are therefore situations in which it can break down, and we shall see in Sec.~IV.C that the heat transport in solid p-H$_2$ is one of them.

\subsection{Computational details}

The processes we shall find it important to include in the collision matrix are three-phonon scattering and boundary scattering, both of which are implemented in the open-source software package \texttt{Phono3py}.\cite{Togo2023}

In the present notation, the perturbation theory formula for the contribution from three-phonon processes is\cite{Cepellotti2016}
\begin{equation}
    \tilde{\Omega}^{(3)}_{\lambda\lambda'} = \frac{A^{(3)}_{\lambda\lambda'}}{\sqrt{\bar{n}_{\lambda}(\bar{n}_{\lambda}+1)\bar{n}_{\lambda'}(\bar{n}_{\lambda'}+1)}},
\end{equation}
where the diagonal elements of the symmetric matrix $\boldsymbol{A}^{(3)}$ are
\begin{equation}
    A^{(3)}_{\lambda\lambda} = 
\sum_{\lambda''\lambda'''}\left(P^{\lambda'''}_{\lambda\lambda''}+\frac{1}{2}P^{\lambda}_{\lambda''\lambda'''}\right),
\end{equation}
and the off-diagonal elements that are neglected in the RTA are
\begin{equation}
    A^{(3)}_{\lambda\lambda'} =
\sum_{\lambda''} \left(P^{\lambda''}_{\lambda\lambda'}-P^{\lambda'}_{\lambda\lambda''}-P^{\lambda}_{\lambda'\lambda''}\right).
\end{equation}
Here $P^{\lambda''}_{\lambda\lambda'}$ is the Fermi golden-rule rate for the coalescence event $\lambda+\lambda'\to\lambda''$,
\begin{align}
P^{\lambda''}_{\lambda\lambda'} &=
\frac{2\pi}{\hbar^2}\sum_{\bf G}
\left|F_{\lambda\lambda'-\lambda''}^{(3)}\right|^2
\bar{n}_{\lambda}\bar{n}_{\lambda'}(\bar{n}_{\lambda''}+1)\nonumber\\
&\times \delta(\omega_{\lambda}+\omega_{\lambda'}-\omega_{\lambda''})\delta_{\q+\q'-\q'',{\bf G}},
\end{align}
where $-\lambda''=(-\q'',s'')$, with
\begin{equation}
F_{\lambda\lambda'\lambda''}^{(3)}\equiv \left(\frac{\hbar^{3}}{8\omega_{\lambda}\omega_{\lambda'}\omega_{\lambda''}}\right)^{1/2}\Phi^{(3)}_{\lambda\lambda'\lambda''},
\end{equation}
and
\begin{equation}
\Phi^{(3)}_{\lambda\lambda'\lambda''}
=
\frac{\partial^3 V^{(c)}(\r)}
{\partial u_{\lambda}\partial u_{\lambda'}\partial u_{\lambda''}}
\Biggr|_{\boldsymbol{u}=\boldsymbol{0}}.
\end{equation}
Note that we have deliberately put the CPMF $V^{(c)}(\r)$ in this expression for the third-order force constants rather than the classical potential $V(\r)$, and that since the Kronecker delta in Eq.~(40) gives
\begin{equation}
    \q+\q'-\q''={\bf G},
\end{equation}
the equation applies equally well to both normal (${\bf G}={\bf 0})$ and Umklapp (${\bf G}\not={\bf 0}$) processes.

The other contribution we shall need to include in the collision matrix is boundary scattering. The three-phonon collision matrix describes the intrinsic scattering within a translationally invariant crystal, whereas boundary scattering is an extrinsic process that breaks the translational symmetry. This relaxes the phonon crystal momentum that is conserved by normal processes and hence limits their long-lived collective transport. \texttt{Phono3py} treats boundary scattering by adding a diagonal scattering term to ${\tilde{\Omega}}$ in the phonon basis to give 
\begin{equation}
\tilde{\Omega}_{\lambda\lambda'}=\tilde{\Omega}^{(3)}_{\lambda\lambda'}+\tilde{\Omega}^{({\rm B})}_{\lambda\lambda'},
\end{equation} 
where
\begin{equation}
\tilde{\Omega}^{({\rm B})}_{\lambda\lambda'} = \frac{|\v_{\lambda}|}{L}\delta_{\lambda\lambda'}
\end{equation}
in which $L$ is a phenomenological mean free path. This term has the effect of lifting the zero eigenvalues in the intrinsic three-phonon collision matrix and regularising the thermal conductivity generated by other long-lived relaxons at low temperatures.

Since experimental samples of p-H$_2$ invariably contain traces of o-H$_2$, there will also be a contribution from impurity scattering. However, if we were to treat this in the same phenomenological way that \texttt{Phono3py} treats boundary scattering, we could simply combine the two processes and regard the resulting $L=L_{\rm B}L_{\rm O}/(L_{\rm B}+L_{\rm O})$ as an effective mean free path for their combined effect. Hence the $L$ in Eq.~(45) should be regarded as an overall mean free path for extrinsic resistive processes that does not contain enough information to differentiate between them. We can nevertheless use this $L$ as a free parameter to fit a thermal conductivity experiment on a sample with a small measured o-H$_2$ concentration and interpret the result as a lower bound on the boundary mean free path $L_{\rm B}$, so that is what we used \texttt{Phono3py} to do. 

 We began by calculating the third-order force constants in Eq.~(42) by finite differences, using the same parameters we used to calculate the second-order force constants in Sec.~III.B (a finite displacement of $0.01\,\mathring{\rm A}$ in a $6\times 6\times 3$ hcp supercell with a force cutoff distance of $9\,\mathring{\rm A}$). We then calculated the collision matrix in Eq.~(37) on a $29\times 29\times 19$ hcp $\q$-mesh, with the energy-conserving Dirac delta functions in Eq.~(40) evaluated using the tetrahedron method.\cite{Blochl1994} We did not have any issues with this method at and above 5 K, but at lower temperatures the collision matrix developed spurious negative eigenvalues indicating a breakdown of the tetrahedron construction. We therefore switched to Gaussian pre-limit delta functions below 5 K, with an overall \texttt{Phono3py} Gaussian width parameter of $\sigma=5$ GHz that was chosen to give good agreement with the thermal conductivity obtained using the tetrahedron method at 5 K and above. The Gaussian functions were found to preserve the positive semi-definiteness of the collision matrix down to 2.5 K. We completed the calculations by using \texttt{Phono3py} to compute the thermal conductivity on a grid of temperatures for various values of the extrinsic mean free path $L$ in Eq.~(45).

\subsection{Results and discussion}

The resulting thermal conductivities for solid p-H$_2$ between 2.5 and 11 K are compared with the experimental measurements of Hill and Schneidmesser\cite{Hill1958} in Fig.~5. We have included BTE results for three values of $L$ and RTA results for the value ($L=0.3$ mm) at which the BTE calculation gives the best match to the experimental data. 

A comparison between the BTE and RTA results with $L=0.3$ mm shows that the RTA is woefully inadequate for solid p-H$_2$ at low temperatures (note the logarithmic scale on the $y$ axis of Fig. 5). At the lowest temperatures in the figure, only the acoustic phonons near the $\Gamma$ point have any appreciable population, so Umklapp processes are suppressed and normal phonon scattering dominates. The RTA discards the off-diagonal normal-scattering contributions to the collision matrix that give rise to long-lived relaxons, and instead treats the diagonal scattering of each phonon mode as independently relaxing its occupation towards thermal equilibrium. This has the wrong physics in the normal regime because normal processes conserve the total phonon crystal momentum. Rather than relaxing the phonons to the equilibrium Bose–Einstein distribution, they establish a collective, heat-carrying drift distribution\cite{Klemens1956}
\begin{equation}
n_{\lambda}^d = \left[e^{(\hbar\omega_{\lambda}-\hbar\q\cdot\boldsymbol{u})/k_{\rm B}T}-1\right]^{-1},
\end{equation}
where $\boldsymbol{u}$ is the collective drift velocity of the phonon gas. As the temperature increases, Umklapp scattering becomes more important and relaxes the phonon momentum, so this collective picture begins to break down. The RTA therefore becomes better justified, as can be seen from the results at the highest temperatures in Fig. 5 (note again the logarithmic scale on the $y$ axis of the figure).

\begin{figure}[t]
    \centering
    \resizebox{0.85\columnwidth}{!}{\includegraphics{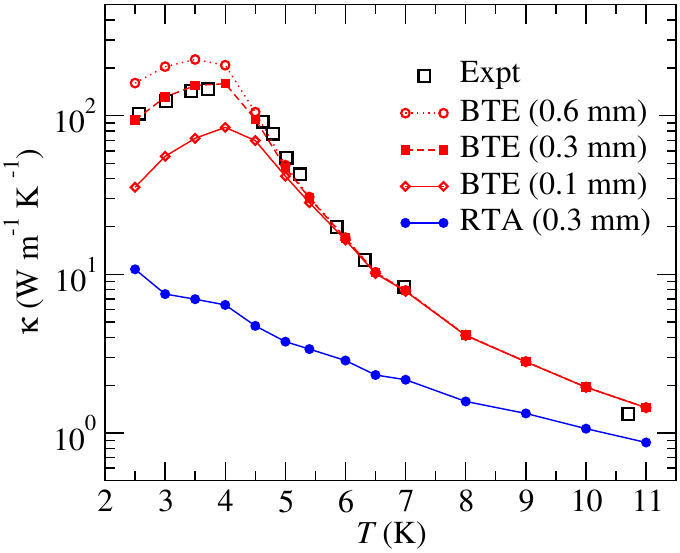}}  
    
    \caption{Comparison of the calculated lattice thermal conductivity $\kappa={\rm tr}[\k]/3$ of solid p-H$_2$ with the experimental results of Hill and Schneidmesser.\cite{Hill1958} The experiment was performed on a sample containing 0.5\% o-H$_2$ and the numbers in parentheses in the captions are the extrinsic mean free paths $L=L_{\rm B}L_{\rm O}/(L_{\rm B}+L_{\rm O})$ used in the calculations.}
    
    \label{fig:5}
\end{figure}

Hill and Schneidmesser used a polycrystalline sample of p-H$_2$ containing 0.5\% o-H$_2$ in the experiment we are comparing with here.\cite{Hill1958} They used the simple phonon kinetic theory formula $\kappa = cvL/3$ to estimate an extrinsic mean free path $L$ from the low-temperature region of their data, where $c$ is the specific heat per unit volume and $v$ is the velocity of sound in the crystal, finding $L\approx 0.3$ mm. They also noted that this contains contributions from both crystallite boundary scattering and o-H$_2$ impurity scattering, and suggested that it was likely to be a considerable under-estimate of the boundary mean free path. However, they were unable to confirm this on the basis of their existing data and argued that an experiment at 1 K would be needed to reduce the remaining impurity scattering enough to do so.\cite{Hill1958}

We find it quite remarkable that the present microscopic intrinsic scattering calculations also give $L=0.3$ mm as the best fit to the experimental data, as can be seen from the BTE curves in Fig.~5. The curve with this extrinsic mean free path is in quantitative agreement with experiment all the way from 2.5 K where the extrinsic scattering has regularised the thermal conductivity from normal processes to 11 K where both normal and Umklapp processes are in play and the effect of the extrinsic scattering has become negligible. It is also interesting to note that this agreement is achieved with a single, temperature-independent $L$. Since the mean free path from crystallite boundary scattering will be temperature independent but the effective mean free path from o-H$_2$ scattering will not, this suggests that the mean free path is dominated by boundary scattering with very little contribution from the o-H$_2$ impurities (i.e., that $L\approx L_{\rm B}\ll L_{\rm O}$). This seems quite plausible given the low concentration of o-H$_2$ in the experiment (0.5\%). However, a more detailed microscopic treatment of the o-H$_2$ impurity scattering\cite{Ebner1970,Kokshenev1975} would be needed to confirm this, and that is not available in the present version of \texttt{Phono3py}.

\section{Conclusion}

The combination of imaginary time path integral and lattice dynamics methods we have used in this paper provides both quantitative agreement with experimental measurements and a convincing explanation for the observed lattice thermal conductivity of solid p-H$_2$ (see Fig.~5). In particular, the calculation shows that the low-temperature heat transport is dominated by collective relaxon modes arising from normal phonon scattering, which is a hydrodynamic process that cannot be described within the phonon RTA.

Lattice dynamics on the classical interaction potential would have failed to reach agreement with experiment because the classical phonon dispersion neglects nuclear quantum effects (see Fig.~3). A Green--Kubo calculation of the thermal conductivity using a path integral-based dynamical method such as CMD\cite{Cao1994b} or ring polymer MD\cite{Craig2004} would also fail at these low temperatures. These methods are poorly justified for calculating the autocorrelation function of the energy flux operator because the classical molecular dynamics they rely on is fundamentally inconsistent with Bose--Einstein phonon statistics. Path integrals are, however, useful for generating the CPMF (see Fig.~1), on which a renormalised BTE lattice dynamics calculation with correctly quantised phonon occupation numbers provides a reliable way to calculate the thermal conductivity.

The success of the present calculations suggests that the same combination of methods could be used to calculate the thermal conductivities of other quantum crystals. For simple monatomic crystals in which the interactions are well described by a pair potential this could be done exactly as we have described in this paper. However, for more interesting crystals, one would need to extract a many-body CPMF from the path integral calculation. The rapid recent development of machine-learning potentials should facilitate this, along with the fact that the CPMF only need be represented accurately in the vicinity of its equilibrium geometry to obtain the force constants required for lattice dynamics.

For solids such as ice with intramolecular bending degrees of freedom there is an additional complication. The optical phonons on the CPMF associated with high-frequency intramolecular stretching vibrations will be artificially red-shifted as a result of their coupling to the bending modes. This phenomenon is well known in the chemical physics literature as the ``curvature problem'' of CMD.\cite{Witt2009} It is possible to fix the problem by replacing the CPMF with a ``quasi-centroid" potential of mean force constructed using appropriate curvilinear coordinates.\cite{Trenins2019} However, that requires a great deal more work, and since the high-frequency phonons have an exponentially small impact on the thermal conductivity at low temperatures in any case, one might be able to get away with simply ignoring the problem.

\begin{acknowledgments}
Z.Z. acknowledges support from a Newton International Fellowship from The Royal Society (No.~NIF/R1/254124). 
\end{acknowledgments}

\section*{Data Availability}

The data that support the findings of this study are available within the article itself. 

\makeatletter
\let\selectlanguage\@gobble
\makeatother

\bibliography{ph2refs}

\end{document}